# Early Insights into System Impacts of Smart Local Energy Systems

M. Aunedi and T.C. Green, Imperial College London

## Summary of Main Findings

A whole-system, investment-optimising model has been used to examine the change in total cost of meeting demand for electricity when Smart Local Energy Systems (SLES) are deployed. Our assumption is that SLES, alongside their other features, enhance the flexibility of electricity consumption through demand-side response (DSR) and facilitate use of local energy storage. We find that:

- With the flexibility of SLES present, variable renewables such as offshore wind can displace firm but more expensive low-carbon sources such as CCS.
- Considering a 100 g$CO_2$/kWh emissions target in 2030, a 10% penetration of SLES could reduce total costs by £1.2bn/year relative to no SLES.
- At higher penetration of 50% SLES, savings increase twofold to £2.8bn/year.
- Under a more stringent emissions limit of 25 g$CO_2$/kWh in 2040, the savings rise to £2.9bn/year for 10% SLES uptake and rise threefold to £8.7bn/year at 50% uptake.
- These results hold for costs of enabling DSR of less than £100/kW and it is not until an unlikely £5,000/kW that the savings are nullified.
- The savings from substituting wind for CCS remain substantial even if the anticipated reduction of cost of wind in 2040 does not materialise.
- Cost savings from the flexibility provided by SLES are affected by realisation of domestic DSR through other means. A 20% uptake level of non-SLES DSR in 2040 still allows SLES to create cost savings of £6.8bn/year at 50% penetration (a 20% fall from £8.7bn/year).

## Introduction

UK's energy sector is expected to undergo a fundamental transformation over the next few decades to deliver on the ambitious target set by the Government to bring all greenhouse gas emissions to net zero by 2050. Decarbonisation of the energy system will require significant and continued investment in low-carbon energy sources such as renewables, nuclear and carbon capture and storage (CCS), and will most likely entail a significant degree of electrification of heat and transport sectors. Although a large fraction of future electricity will be provided through investment in large-scale low-carbon technologies (e.g. offshore wind) and will therefore flow from the transmission network to the end users, the provision of flexibility and resilience is thought likely to shift towards decentralised and distributed sources provided by consumers of end-use energy [1]. Recent analysis [2] has demonstrated that cost savings in operating and investment costs from the application of flexible technologies could reach £8bn/year in 2030. This arises because flexibility, such as in terms of when energy is used, enables energy needs to be satisfied with a smaller investment in generation and network infrastructure and then making use of that generation fully whenever it is able to provide.

There are, however, a number of barriers to using decentralised assets to provide flexibility under the current paradigm of centralised system design and operation. Smart Local Energy Systems (SLES) are seen as a vehicle for unlocking the potential for decentralised flexibility, driven not only by an increased general recognition of the importance of flexibility but also by local stakeholders seeking to align the development of local energy systems with the objectives of the local community. There is currently a lack of detailed knowledge as to the magnitude and nature of the contribution that SLES can make to not just local but also national objectives such as cost-efficient decarbonisation and quite what economic costs and benefits arise.

### *Potential benefits of SLES*

There is no single definition of SLES as there is a variety of ways in which they could be implemented and configured, including variations in geographical boundaries, energy vectors included, types of assets (generation, storage, flexible demand) and the actors involved. Analytical approaches have been proposed in the literature to evaluate local energy systems on a wide scale and it is common to reduce the complexity of the problem by specifying a manageable and representative number of archetypal local energy systems [3].

A framework for understanding the process by which SLES could potentially deliver both system and societal benefits has been established in [4]. One of the key aspects of the system value of SLES is the set of benefits arising from providing local flexibility to the wider energy system, thus contributing to a more cost-effective integration of renewable energy sources (RES) [5]. A recent report [6] stressed the need for end-use customers to become engaged at a local level with the transformation to a decarbonised and less-centralised energy system, through changes in individual homes (energy efficiency measures and new heating technologies), cars (electric vehicles) and types of household appliance (providing options for demand response). The report estimates that with the right policy support, the community energy sector in the UK could grow to between 12 and 20 times larger than today by 2030 and could encompass up to 4,000 organisations.

### *Objective*

The objective of this *Briefing Note* is to assess the value to the whole energy system of SLES, with a particular focus on the whole-system benefits of local flexibility resources that might be unlocked and enhanced by SLES. For the purpose of the analysis presented here, flexibility is assumed to entail any action that is taken in response to system conditions by customers such as modifying their original energy use profiles or moving their energy use in time, or actions taken by energy storage assets to inject or absorb energy from the grid. The assessment is carried out against the backdrop of a transition to a low- or zero-carbon energy system in the 2030-2050 horizon.

## Modelling approach

In order to assess the benefits of SLES for the UK's future low- or zero-carbon electricity system, we have modelled the flexibility released through SLES and incorporated it into our whole-system modelling framework, allowing us to directly quantify cost savings arising from SLES deployment. More detail on our whole-system modelling approach is provided in Box 1.

### *Including SLES in whole-system modelling*

Developing a generalised representation of SLES in an energy system model such as WeSIM is not straightforward given the large diversity of possible SLES configurations. Nevertheless, for the purpose of this note, it was assumed that SLES have the potential to unlock flexibility resources at or close to end-use customers. Evidence to support this assumption is discussed in Box 2. More specifically, SLES are assumed to facilitate uptake of very small-scale DSR and highly distributed battery storage (at kilowatt scale) much more so than with current market signals alone. These small-scale DSR and battery assets would be connected at the low-voltage level of the local distribution grids, thus creating opportunities to deliver both highly localised benefits and grid services to the national transmission system. It was assumed that the flexible loads able to provide DSR services which are enabled by SLES will include: electric vehicles, electric and hybrid heat pumps and smart domestic appliances. Thus, by stating a certain penetration of SLES there is an assumed enablement of a proportional share of highly distributed DSR and battery storage.

> **Box 1. Whole-system modelling of low-carbon electricity systems**
>
> Capturing the interactions across various time-scales and across different types of asset at sufficient temporal and spatial granularity is essential for the analysis of future low-carbon electricity systems in order to take proper account of flexible technologies such as energy storage and demand-side response (DSR). The Whole-electricity System Investment Model (WeSIM) has been developed at Imperial College London (under the leadership of Prof Goran Strbac) in order to capture the effects and trade-offs between different flexible technologies [7].
>
> WeSIM is a comprehensive system analysis model that is able to simultaneously optimise decisions on long-term investment into generation, network and storage assets, and short-term operation decisions in order to satisfy the defined electricity demand at least cost, while also ensuring adequate security of supply and sufficient volumes of ancillary services. The optimisation can be constrained to meet a carbon emission target.
>
> WeSIM is capable of quantifying trade-offs between using various sources of flexibility, such as DSR and energy storage, for real-time balancing and for management of transmission and distribution network constraints, thus capturing the synergies and conflicts between local/district-level and national-level infrastructure requirements. A prominent feature of WeSIM is the ability to quantify the necessary investments in distribution networks in order to meet demand growth and/or distributed generation uptake, based on the concept of statistically representative distribution networks.

> **Box 2. Previous evidence on drivers for unlocking local flexibility potential**
>
> A key driver for the value of SLES quantified in this analysis is the argument that SLES would drive a higher uptake of local energy storage and encourage greater participation in demand-side response (DSR). A recent briefing paper by Imperial College's Energy Futures Lab [9] comprehensively assessed the evidence base on residential consumer engagement with DSR to identify barriers, drivers and opportunities for greater household consumer engagement. One of the findings was that DSR engagement is higher if consumers see a link to maximising the use of renewable energy and higher when there is a combination of supportive technologies such as EV, storage and smart appliances and smart heating controls. Therefore if a SLES combines these factors and achieves closer engagement with customers such as promoting understanding of energy issues, encouraging deployment of on-site renewables (PV) and storage and supporting adoption of EVs and smart appliances and heating, it is plausible to assume that SLES will indeed increase participation in DSR.
>
> ClientEarth's annual survey of UK attitudes towards climate change clearly suggests that the majority of consumers would like to install both solar panels and a home energy storage device for their homes, or switch to an electric or low-carbon vehicle, if greater assistance was available from the UK government or through community or commercial schemes [10].
>
> The study by Fell et al. on the effect of tariff design and marketing on willingness of consumers to adopt time-of-use and demand-side response tariffs [11] concluded that neither age, gender, housing tenure, employment status, education, social grade, being on a pre-payment meter, nor income were consistently associated with being more or less willing to switch tariffs. On the other hand, trust in suppliers was the most important predictor of using DSR tariffs and services, as people who trust their electricity supplier were more likely to say they would switch to a DSR tariff, while people who were concerned about their privacy were less likely to. This supports the assumption that a SLES with a close and trusted relationship with the customers would achieve enhanced DSR engagement.

*System scenarios and cases for SLES uptake*

Analysis has been conducted for two broad time horizons, 2030 and 2040, based on power system scenarios defined for a recent CCC study [8]. The associated carbon targets, minimum RES portfolios and the projected uptakes of electrified transport and heating are provided in Table 1. The carbon intensity targets of 100 and 25 g$CO_2$/kWh refer to the carbon footprint of GB electricity generation and are quantified as total annual $CO_2$ emissions from electricity generation divided with total annual generation output.

In all cases, the model was allowed to add more offshore wind, solar PV and CCS capacity (with new CCS in 2030 limited to 1.5 GW) in order to meet the carbon target at the lowest cost. Note that any investment in additional solar PV generation is proposed by the optimiser only if it is cost-optimal for the overall system and in particular it is not assumed that the PV deployment is directly a feature of SLES deployment. The levelised cost of offshore wind was assumed to be £40/MWh in 2030 and £35/MWh in 2040, for PV generation the assumptions were £55/MWh in 2030 and £50/MWh in 2040, while the levelised cost of CCS (assuming 90% annual load factor and operating at full output) was £94/MWh. A separate sensitivity study was carried out with a higher cost of offshore wind, equal to the LCOE of solar PV.

*Table 1. Assumptions for electricity system scenarios*

| Time horizon | Carbon target (g$CO_2$/kWh) | Generation capacity (GW) | Electrified transport and heat (million units)* |
|---|---|---|---|
| 2030 | 100 | Wind: 47.5, PV: 30.9, Nuclear: 4.5, CCS: 0.0 | EV: 12.2, HP: 2.2, HHP: 0.0 |
| 2040 | 25 | Wind: 58.1, PV: 37.0, Nuclear: 7.9, CCS: 2.0 | EV: 37.1, HP: 5.7, HHP: 9.5 |

* EV = Electric Vehicles, HP = Heat Pumps, HHP = Hybrid Heat Pumps

On top of the two system scenarios, four SLES uptake cases were considered in each time horizon: Counterfactual (no SLES), Low SLES (10% uptake), Medium SLES (25% uptake) and High SLES (50% uptake). The uptake of SLES is defined here as the share of electricity demand that is supplied through SLES. The assumptions on deployment of DSR and battery storage that occurs as a part of SLES uptake are set out in Table 2.

*Table 2. Assumptions for SLES uptake cases*

| SLES case | SLES & LV DSR uptake | 2030 | | | 2040 | | |
|---|---|---|---|---|---|---|---|
| | | HV DSR uptake | HV storage (GW) | LV storage (GW) | HV DSR uptake | HV storage (GW) | LV storage (GW) |
| Counterfactual | 0% | 25% | 7 | 0 | 50% | 10 | 0 |
| Low SLES | 10% | 25% | 7 | 2 | 50% | 10 | 5 |
| Medium SLES | 25% | 25% | 7 | 5 | 50% | 10 | 10 |
| High SLES | 50% | 25% | 7 | 10 | 50% | 10 | 20 |

The uptake of high-voltage (HV) DSR provided by large-scale industrial and commercial (I&C) consumers is assumed to be independent of SLES uptake, and, further, assumed to increase over time to 25% in 2030 and 50% in 2040 in line with decarbonisation and RES integration objectives. Note that the assumption on the DSR uptake in the context of WeSIM refers to the volume of DSR relative to its maximum theoretical potential, which is quantified separately for each demand category (I&C demand, electric vehicles, heat pumps etc.) based on the authors' previous bottom-up modelling of demand flexibility. This flexibility assessment took into account customer requirements to ensure there is no compromise on the level of service delivered (e.g., all vehicles will be charged when needed and home temperature maintained within comfortable limits).

For example, for a full DSR penetration (i.e., 100% uptake) it is assumed that up to 10% of I&C demand can be shifted away from peak hours to other hours in the same day. For a lower DSR uptake level, e.g. 25%, this assumption is scaled down proportionally, so that 25% uptake allows up to 2.5% of the baseline demand to be shifted in each 24-hour period. The uptake of low-voltage (LV) DSR on the other hand was assumed to be directly proportional and equal in magnitude to the SLES uptake, reflecting the assumption that SLES will be the key means to deliver highly distributed flexibility. Nevertheless, we also explored the sensitivity of our findings to assuming a non-zero residential DSR uptake in Counterfactual.

In all cases the model was allowed to add HV battery storage (at megawatt-scale) with volumes of up to 7 GW in 2030 and 10 GW in 2040 at a cost of £180/kWh. Deployment of large-scale HV-connected battery projects is assumed to be largely independent from SLES uptake, as reflected in the current typical sizes of battery projects built by commercial developers. The LV battery storage on the other hand was assumed to be driven by SLES, and the maximum volumes broadly proportional to SLES uptake. The cost of LV battery systems was assumed to be 50% higher than for HV battery storage to reflect economies of scale in balance-of-plant and construction. Note that the battery volumes in Table 2 reflect the upper limits of what the model is allowed to build; the actual cost-optimal volume may be lower than that.

A limitation at this stage is that the investment cost of implementing or enabling DSR schemes is highly uncertain and so it has not been included in the main set of results from the analysis. The expectation is that the Prospering from the Energy Revolution (PFER) demonstrators will provide some data on cost per customer or cost per unit power as the trials progress.

## Key insights

The focus of the quantitative studies of benefits of SLES presented here is to explore how the distributed flexibility unlocked and enhanced through SLES can contribute, potentially, to more cost-efficient operation and investment in the future low-carbon electricity system, including:

- Reduced total system cost,
- Lower levels of peak electricity demand,
- Reduced cost of investing in (and operating) low-carbon generation to meet a given carbon target,
- Avoidance of reinforcement of local distribution networks.

This section presents the key numerical outputs of the model that quantify the benefits of SLES across various SLES and system scenarios (and under the assumptions stated).

### System value of SLES

The value of SLES is quantified here as the difference in total system cost between a given SLES uptake case and the relevant counterfactual case for a given system scenario. Cost-benefits of SLES are disaggregated into components referring to investment cost (CAPEX) for generation, network and storage assets, as well as operating cost (OPEX) of electricity generation. Note that the benefits of SLES represent net system value in the sense that they include the investment cost of LV battery storage as negative component of the benefit but, as noted above, the cost of implementing DSR schemes is not included in the main results, although it is included in a sensitivity analysis later in the document.

Net system benefits of SLES against relevant counterfactual (i.e., no-SLES) cases in 2030 and 2040 are shown in Figure 1. The cost advantages are broken-down into the following components:

- Generation CAPEX (low-C): investment cost in low-carbon generation (RES, nuclear and CCS)
- Generation CAPEX (other): investment cost in conventional gas generation (CCGT and OCGT)

- Interconnection / transmission / distribution CAPEX: investment cost in network infrastructure
- Storage CAPEX: investment cost in energy storage assets
- OPEX (low-C): operating cost of low-carbon generation (CCS and nuclear)
- OPEX (other): operating cost of gas generation (CCGT and OCGT)

Note that while most of these components are positive, that is, the SLES case has lower cost than no-SLES case, some components can also take negative values. Notably the storage CAPEX advantage is negative because it is assumed that SLES uptake drives additional LV battery storage investment compared to the Counterfactual case. Furthermore, OPEX (other) advantage can be negative when some of the low-carbon generation output from, for instance, CCS plants is replaced by less expensive (but also more carbon-intensive) unabated gas generation.

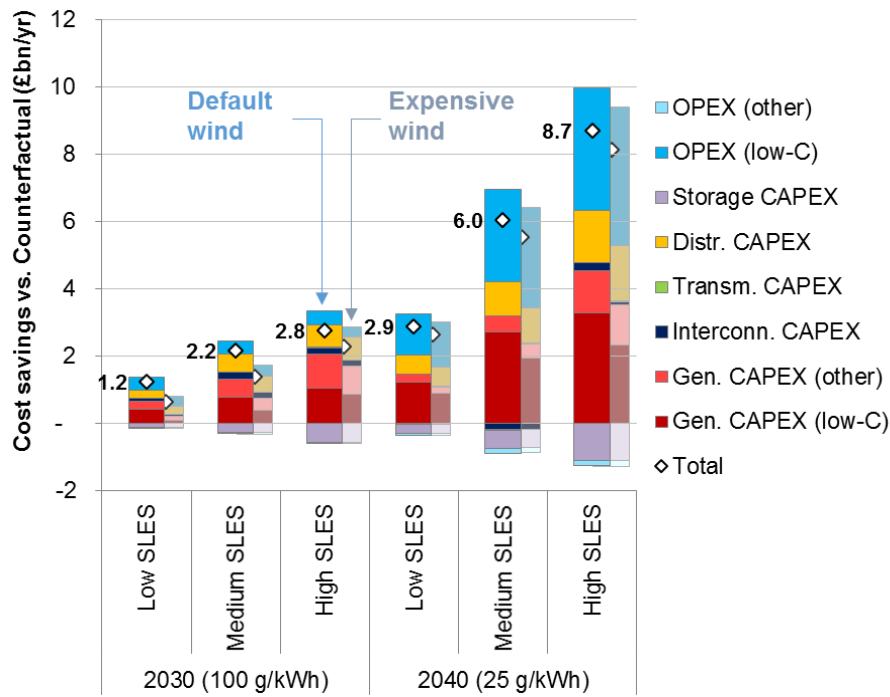

*Figure 1. Net system benefits of SLES*

The first remark is that all SLES deployments, on both time horizons, show a saving in total system cost compared to the no-SLES counterfactual. This suggests that the flexibility that SLES are assumed to bring (through enhancing DSR or facilitating local storage) does indeed lead to cost savings elsewhere in the wider system. The results also show that much higher cost savings are achieved in the 2040 case where the flexibility offered by SLES enables large cost savings in the CAPEX and OPEX of the low carbon generation that is needed to meet the lower carbon emissions target of 25 g/kWh. For instance, in the Medium SLES case the net benefits increase almost 3 times, from £2.2bn per year in 2030 to £6.0bn per year in 2040. SLES benefits also increase at higher SLES uptake levels, although not in direct proportion: increasing the SLES uptake in 2040 by 2.5 times (from Low to Medium) and 5 times (Low to High) increases the benefits by a factors of 2.1 and 3.0, respectively. This suggests that benefits begin to saturate, where the early additions of flexible resources deliver the highest benefits whereas further additions result in diminishing contributions to cost savings. Nevertheless, the benefits of SLES in 2040 are substantial, reaching £8.7bn per year in the High SLES case, which is broadly similar to the £8bn per year saving attributed to smart flexible systems in [2].

There are several key components in the breakdown of SLES-driven system benefits. As stated before, increased investment in LV battery storage driven by SLES is reflected in a negative storage CAPEX benefit for all cases, which is, however, far smaller than the positive benefit components. A significant component of the benefits is represented by avoided distribution CAPEX, driven by reduced net loading of distribution grids as the result of using highly distributed DSR and battery storage resources. The dominant components of system benefits are driven by reductions in generation investment and operation cost, in particular through significantly lower investment in low-carbon generation.

An additional set of case studies was carried out to test the sensitivity of system benefits of SLES to the assumption on the future cost of offshore wind generation, i.e., to analyse how the findings might change if offshore wind does not turn out to benefit from further cost reductions. The results of the sensitivity analysis are also shown in Figure 1 using lower-intensity colour bars in the background labelled "Expensive wind", suggesting that a higher cost of wind would effectively reduce the system value of SLES. This can be explained by the lower cost differential between the CCS, which the optimiser uses extensively in the Counterfactual case, and the wind generation that displaces it as SLES uptake increases. SLES still facilitates the uptake of wind but the economic benefit of that is smaller. Nonetheless, there is still significant system benefit of SLES, especially in the 2040 horizon, and the sensitivity to the cost of offshore wind is relatively small. Clearly there are further sensitivities yet to explore such as steeper cost reductions in wind and variations for storage and PV costs.

*Sensitivity to assumptions on DSR*

In order to mitigate the high uncertainty of the cost of DSR and its impact on the net benefits of SLES, we explored various assumptions for DSR cost. The cost of DSR was expressed in monetary units per unit of flexible power offered by residential DSR, and was varied between £100/kW and £5,000/kW (note that, for instance, a recent report [12] quoted the cost estimates for residential DSR in the range between £23/kW and £805/kW and note also that Figure 1 already covered the £0/kW case). The net saving in system cost due to SLES (savings as found in main results of Figure 1 with the cost DSR deducted) are presented in Figure 2a alongside gross savings with no DSR cost. The results for net savings clearly suggest that it is not until DSR costs reach several hundred pounds per kilowatt that a significant negative impact on net benefits of SLES is found. Furthermore, the cost savings of SLES are not nullified until the cost of enabling DSR reaches several thousand pounds per kilowatt of DSR capacity and such a cost is at or above the upper limit of the range of DSR cost estimates available. It is reasonable to say that the benefits of SLES quantified in the main analysis are robust to variation of the cost of enabling DSR.

The nature of the counterfactual is important to understand in that it is pessimistic over uptake of demand-side response (DSR), for instance there is a large deployment of EV in 2040 but they are not used for DSR. This means that SLES can capture value from the first tranche of DSR whereas in a more optimistic counterfactual some of that value would be captured by DSR outside the SLES. In order to explore the impact of a higher residential DSR uptake in the Counterfactual on the value of SLES, we carried out another set of sensitivity studies with modified assumptions on the counterfactual uptake of residential DSR. The alternative DSR uptake for smart residential EV, appliance and heating demand for this sensitivity was assumed to be 10% in 2030 and 20% in 2040 in the no-SLES case. The results for this alternative counterfactual are presented in Figure 2b alongside previously presented results from the main studies.

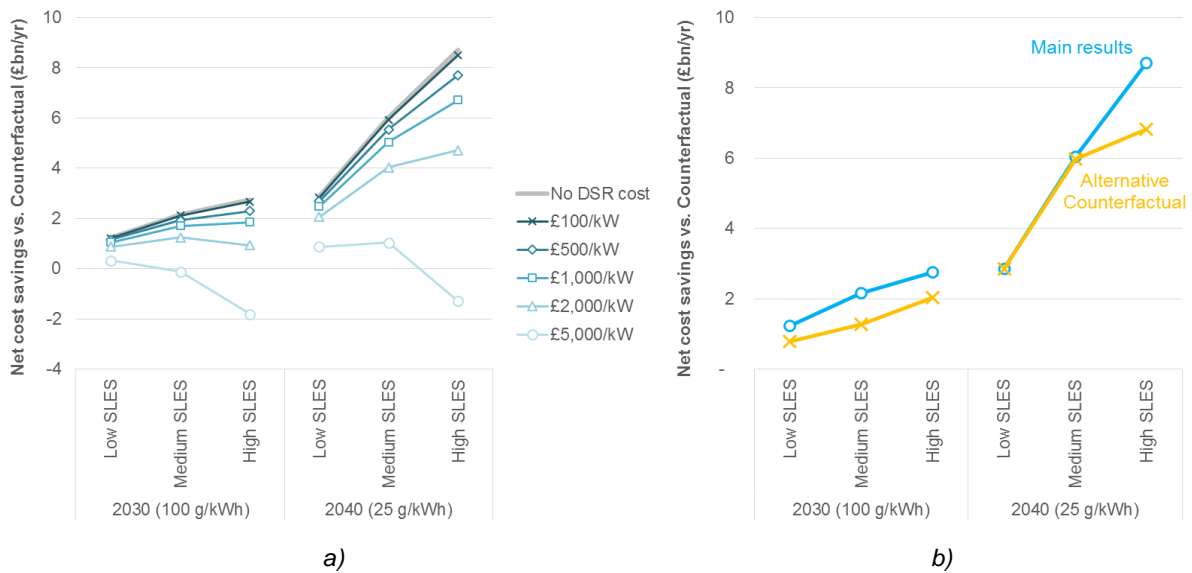

*Figure 2. Sensitivity of net system benefits of SLES to: a) DSR cost, and b) residential DSR uptake in Counterfactual case*

In 2030 we observe a decline in the value of SLES (across various SLES uptake levels) to 25-40% below the benefits found in the main case studies. In 2040, on the other hand, the value of SLES is unaffected by alternative counterfactual assumptions for Low and Medium SLES uptake cases, but the value falls in the High SLES uptake case by about 20% compared to main results. This can be explained by changes in the 2040 low-carbon generation portfolio driven by SLES, where offshore wind displaces CCS as more flexibility is made available in the system. Unlike in the original case studies, where CCS is displaced by offshore wind at all SLES uptake levels, with alternative counterfactual, which contains less CCS because of its DSR, all of the added CSS is already displaced at the Medium SLES uptake level. Therefore, the incremental benefit of increasing SLES uptake from Medium to High is lower and arises only from displacing peaking generation capacity and avoiding distribution network reinforcement because the value of replacing CCS with wind has already been exhausted. Nevertheless, even with more optimistic counterfactual assumptions on residential DSR, the system value of SLES for this case is still a very significant £6.8bn per year.

### *Key drivers for system benefits of SLES*

To provide further insights on key drivers behind generation CAPEX and OPEX savings presented in Figure 1, it is useful to examine how increasing SLES uptake changes the cost-optimal generation capacity mix and changes the annual energy output of the various technologies. Figure 3 shows how progressively increasing SLES uptake affects the cost-optimal generation mix (note that Figure 3 shows the results for default assumptions on offshore wind cost and is therefore consistent with the net benefits shown in Figure 1). As a first observation, the optimiser decides to add all available LV battery storage capacity in each SLES case, according to limits specified in Table 2. The additional capacity of battery storage and DSR unlocked by SLES reduces the need for firm generation capacity, i.e., gas-fired OCGT and CCGT plants, which drives the saving in the generation CAPEX (other) in Figure 1.

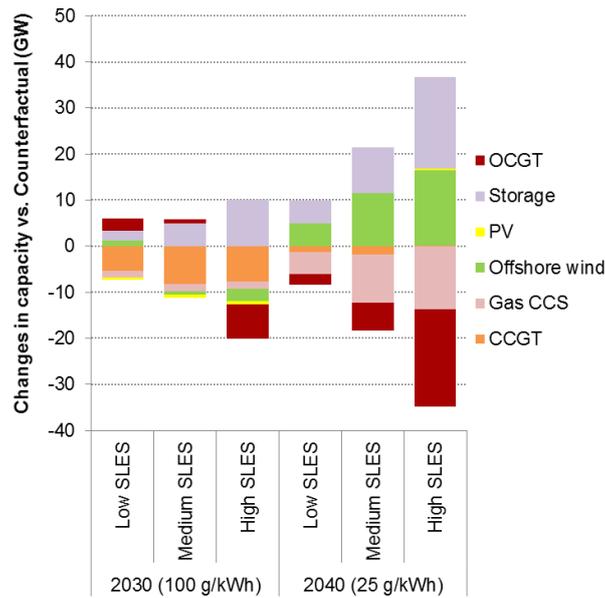

*Figure 3. Changes in installed capacity vs. Counterfactual driven by SLES*

Savings in low-carbon generation CAPEX, on the other hand, result from the changes in the cost-optimal portfolio of low-carbon generation. In 2030 the additional flexibility facilitated through SLES effectively reduces the total CAPEX of low-carbon generation by displacing 1.5 GW of CCS capacity and adjusting the offshore wind and PV capacity required to meet the carbon target. In 2040, offshore wind replaces CCS capacity, delivering overall low-carbon CAPEX savings (note that the assumed investment cost per MW of CCS capacity is 3.1 times higher than for offshore wind and 7.6 times higher than for PV).

Changes in annual generation output compared to the counterfactual, presented in Figure 4, explain what lies behind the OPEX components of SLES benefits. The prominent low-carbon OPEX savings component in 2040 results from wind output, assumed to be available at zero marginal cost, displacing CCS output with the variable operating cost of around £45/MWh. Unlike wind and PV, gas-fired CCS generation is not fully zero-carbon, as it typically emits about 10% of the non-abated carbon emissions from a similar-size CCGT plant. As a consequence, when wind (or PV) output displace CCS, the system is able to accommodate slightly more CCGT output while maintaining the same level of carbon emissions; this is reflected in a moderate increase in CCGT output alongside wind, as well as in a small negative OPEX (other) component of system benefit in Figure 1.

Finally, it is useful to quantify how the savings delivered by SLES through unlocking and enhancing local flexibility compare to the overall total system cost. To illustrate this, Table 3 expresses the total system cost as the average cost of electricity in £/MWh, calculated by dividing the total annualised system cost by total annual demand. In other words, this cost includes the cost of owning and operating the network and so is higher than the LCOE of the generation technologies. Relative cost reductions delivered by SLES are significant, up to 6% in the Low SLES case and up to 19% in the High SLES case. Higher relative cost reductions are observed in 2040 cases.

*Initial conclusions on impact of SLES on electricity infrastructure requirements*

A factor in the system value of SLES is the reduced need for investing in electricity generation and the local and national network infrastructure. This results from SLES facilitating the use of local DSR and local energy storage resources. The use of DSR and local storage reduce the net peak loading of the power system infrastructure, allowing the demand to be met with less installed generation capacity and less network capacity while maintaining the same level of security of supply. On the generation side, this effect is seen as a reduced need for OCGT capacity for "peaking", as was illustrated in Figure 3.

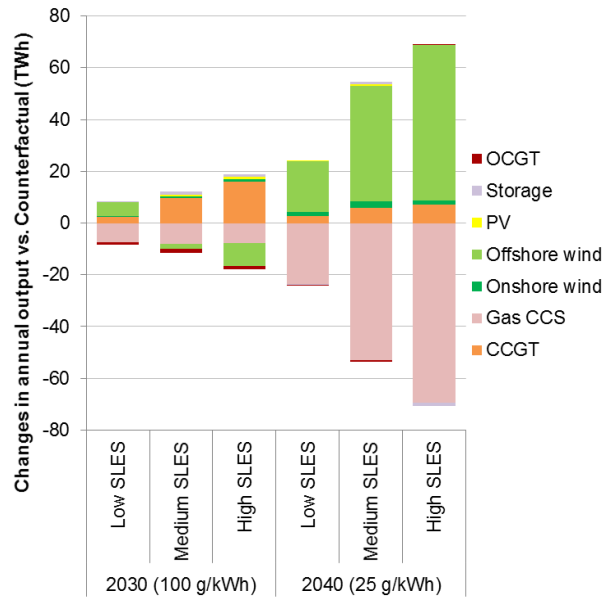

*Figure 4. Changes in annual output vs. Counterfactual driven by SLES*

Table 3. Average cost of electricity across different SLES cases

| SLES case | 2030 (100 gCO$_2$/kWh) | | 2040 (25 gCO$_2$/kWh) | |
|---|---|---|---|---|
| | Avg. cost of electricity (£/MWh) | Reduction vs. Counterfactual | Avg. cost of electricity (£/MWh) | Reduction vs. Counterfactual |
| Counterfactual | 95.4 | - | 103.2 | - |
| Low SLES | 91.8 | -3.8% | 96.6 | -6.4% |
| Medium SLES | 89.0 | -6.7% | 89.2 | -13.6% |
| High SLES | 87.2 | -8.6% | 83.5 | -19.1% |

On the network side, as Figure 5 shows, the net peak system demand after accounting for DSR and storage actions changes across the various SLES uptake cases. Three peak demand levels are shown: a) "Pre-DSR", which is the system peak demand before accounting for any DSR or storage actions, b) "Post-DSR", which accounts for DSR actions only, and c) "Net of BESS & PV", which provides an indication of the actual net loading of the distribution grid after accounting for use of DSR and battery storage plus accounting for on-site generation by PV (although the energy production of solar PV is essentially zero at the time of winter peak demand). The flexibility facilitated through SLES is seen to have a very significant effect on net peak demand: in 2040 the net peak loading of the distribution grid reduces from 80 GW for the Counterfactual to 63 GW for High SLES case. Figure 5 also shows the required level of distribution network reinforcement (as presented earlier in Figure 1), which shows a clear decreasing trend with increasing SLES uptake. As an example, distribution network reinforcement cost in 2040 reduces from £2.7bn per year in the Counterfactual case to £1.1bn per year in the High SLES case.

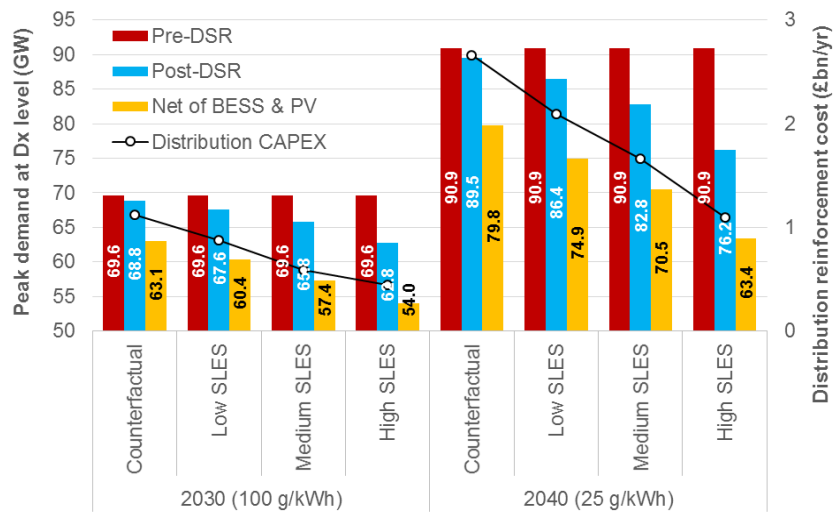

*Figure 5. Net system peak demand and distribution network reinforcement cost*

## Future work

The system benefits of SLES quantified in this note are driven by the underlying assumption that SLES can unlock and enhance flexibility from highly distributed resources such as residential DSR and small-scale battery storage. Clearly it will be necessary to explore how realistic these are assumptions are as insights are gained from PFER demonstrators and from analysis of SLES in other areas. Even so, data will be sparse and the sensitivity of these initial results to variations in the set of cost and deployment assumptions should be undertaken. Costs of several key technologies, notably batteries, PV panels and wind farms have decreased as volumes expand at rates higher than forecast. There is considerable uncertainty over what cost decreases will occur over the next few decades.

We have not as yet explored the differences in system impacts that might arise between concentrated and dispersed versions of SLES in which assets such as storage, PV and DSR are either all located in a relatively narrow geographic area or intermingled with non-SLES assets across a larger area. Further work to characterise the operational differences is needed and then means to represent the differences will be formulated in WeSIM.

The results here take us as far as 25 g/kWh in electricity in 2040 but not to a net-zero national energy system including all energy vectors (electricity, heat, hydrogen and gas). Exploring how vector interactions within a SLES context could further contribute to a cost-efficient operation and design of an integrated low-carbon energy system is an important modelling extension to undertake. It is of particular interest to investigate the potential for SLES to coordinate flexibility provided by district heating systems through controlling power and heat sources and utilising thermal storage and the thermal inertia of pipes and buildings.

Future work will also focus on investigating the self-sufficiency aspects of SLES in more detail, and in particular how targeting various local objectives might change the requirements for transmission and distribution network capacity. Future research will also address how the delivery of local flexibility may be constrained by upstream network capacity and how the associated cost of upgrade may affect the benefit of SLES.

Finally, a critical aspect of future work on whole-system modelling of SLES will be to link this research with other modelling approaches being explored elsewhere in WP5.3, most notably the Cardiff University's EnergyHub model and UCL's BRAIN model. The ambition is to enable the whole-system modelling to reflect more refined considerations of local assets and operating strategies addressed in EnergyHub, as well as to enable a better representation of how different agents make investment decisions in actual energy market environment, which is the main focus of the BRAIN model.

*Authors and affiliations*

M. Aunedi and T.C. Green, Imperial College London

*References*

[1] Strbac, G., Pudjianto, D., Aunedi, M., Papadaskalopoulos, D., Djapic, P., Yujian Ye, Moreira, R., Karimi, H., Ying Fan, "Cost-Effective Decarbonization in a Decentralized Market: The Benefits of Using Flexible Technologies and Resources", *IEEE Power & Energy Magazine*, 2019, 17:25–36.

[2] Imperial College London, "Value of Flexibility in a Decarbonised Grid and System Externalities of Low-Carbon Generation Technologies", Report for the Committee on Climate Change, October 2015. https://www.theccc.org.uk/wp-content/uploads/2015/10/CCC_Externalities_report_Imperial_Final_21Oct20151.pdf

[3] Yazdanie, M., Densing, M., Wokaun, A., "The nationwide characterization and modeling of local energy systems: Quantifying the role of decentralized generation and energy resources in future communities", *Energy Policy*, 2018, 118:516–533.

[4] Ford, R., Maidment, C., Fell, M., Vigurs, C., and Morris, M. 2019. A framework for understanding and conceptualising smart local energy systems. EnergyREV, Strathclyde, UK. University of Strathclyde Publishing, UK. ISBN: 978-1-909522-57-2. https://www.energyrev.org.uk/media/1298/energyrev_paper_a-framework-for-sles_20191018correct.pdf

[5] Thellufsen, J.Z., Lund, H., "Roles of local and national energy systems in the integration of renewable energy", *Applied Energy*, 2016, 183:419–429.

[6] WPI Economics, "The future of community energy", report for SP Energy Networks, January 2020. http://wpieconomics.com/publications/future-community-energy/

[7] Pudjianto, D., Aunedi, M., Djapic, P., Strbac, G., "Whole-system assessment of value of energy storage in low-carbon electricity systems", *IEEE Transactions on Smart Grid*, 2014, 5(2):1098–1109.

[8] Vivid Economics and Imperial College London, "Accelerated electrification and the GB electricity system", report for the Committee on Climate Change, April 2019. https://www.theccc.org.uk/publication/accelerated-electrification-and-the-gb-electricity-system/

[9] Carmichael, R., Gross, R., Rhodes, A., "Unlocking the potential of residential electricity consumer engagement with Demand Response", Energy Futures Lab Briefing Paper, November 2018. https://www.imperial.ac.uk/energy-futures-lab/policy/briefing-papers/paper-3/

[10] ClientEarth's Climate Snapshot 2019. https://www.documents.clientearth.org/wp-content/uploads/library/2019-10-30-clientearth-climate-snapshot-2019-ce-en.pdf

[11] Fell, M.J., Nicolson, M., Huebner, G.M., Shipworth, D., "Is it time? Consumers and time of use tariffs", March 2015. https://www.smartenergygb.org/en/~/media/SmartEnergy/essential-documents/press-resources/Documents/UCL-research-into-time-of-use-tariffs.ashx

[12] Imperial College London and Carbon Trust, "An analysis of electricity system flexibility for Great Britain", report for BEIS, November 2016. https://assets.publishing.service.gov.uk/government/uploads/system/uploads/attachment_data/file/568982/An_analysis_of_electricity_flexibility_for_Great_Britain.pdf

[13] Koirala, B.P., van Oost, E., van der Windt, H., "Community energy storage: A responsible innovation towards a sustainable energy system?", *Applied Energy*, 2018, 231:570–585.